\documentclass[lettersize,journal]{IEEEtran}

\usepackage{amsmath,amsfonts}
\usepackage{amssymb}
\usepackage{algorithmic}
\usepackage{algorithm}
\usepackage{array}
\usepackage[caption=false,font=normalsize,labelfont=sf,textfont=sf]{subfig}
\usepackage{textcomp}
\usepackage{stfloats}
\usepackage{url}
\usepackage{verbatim}
\usepackage{graphicx}
\usepackage{cite}
\usepackage{orcidlink}
\usepackage{dsfont}

\usepackage{multirow}
\usepackage{multicol}
\newcolumntype{d}[1]{D..{#1}}

\usepackage{booktabs}
\usepackage{tcolorbox}
\usepackage{flushend}

\newcommand{\PTrial}{$\checkmark$}
\newcommand{\Penroll}{$\times$}

\newcommand{\Nenroll}{$\times$}

\begin{document}

\title{Joint Speaker Encoder and Neural Back-end Model for Fully End-to-End Automatic Speaker Verification with Multiple Enrollment Utterances}

\author{Chang Zeng \orcidlink{0000-0002-4882-1823}, \IEEEmembership{Student Member, IEEE}, Xiaoxiao Miao \orcidlink{0000-0002-6645-6524}, \IEEEmembership{Member, IEEE}, Xin Wang \orcidlink{0000-0001-8246-0606}, \IEEEmembership{Member, IEEE}, \\
Erica Cooper \orcidlink{0000-0002-2978-2793}, \IEEEmembership{Member, IEEE} and Junichi Yamagishi \orcidlink{0000-0003-2752-3955}, \IEEEmembership{Senior Member, IEEE}

\thanks{Manuscript received XXX.

This study was partially supported by JST CREST Grants (JPMJCR18A6 and JPMJCR20D3), MEXT KAKENHI Grants (21K17775, 21H04906, 21K11951, 18H04112), and Google AI for Japan program.
%
\textit{(Corresponding author: Chang Zeng.)}}
\thanks{Chang Zeng, Xiaoxiao Miao, Xin Wang, Erica Cooper and Junichi Yamagishi are with the National Institute of Informatics, 2-1-2 Hitotsubashi Chiyoda-ku, Tokyo 101-8340, Japan (e-mail: \{zengchang, xiaoxiaomiao, wangxin, ecooper, jyamagis\}@nii.ac.jp).}}

\markboth{Journal of \LaTeX\ Class Files,~Vol.~XX, No.~X, XXX~2022}%
{Shell \MakeLowercase{\textit{et al.}}: The PartialSpoof Database and Countermeasures for the Detection of Short Generated Audio Segments Embedded in a Speech Utterance}

\maketitle

\begin{abstract}
Conventional automatic speaker verification systems can usually be decomposed into a front-end model such as time delay neural network (TDNN) for extracting speaker embeddings and a back-end model such as statistics-based probabilistic linear discriminant analysis (PLDA) or neural network-based neural PLDA (NPLDA) for similarity scoring. However, the sequential optimization of the front-end and back-end models may lead to a local minimum, which theoretically prevents the whole system from achieving the best optimization. Although some methods have been proposed for jointly optimizing the two models, such as the generalized end-to-end (GE2E) model and NPLDA E2E model, all of these methods are designed for use with a single enrollment utterance. In this paper, we propose a new E2E joint method for speaker verification especially designed for the practical case of multiple enrollment utterances. In order to leverage the intra-relationship among multiple enrollment utterances, our model comes equipped with frame-level and utterance-level attention mechanisms. We also utilize several data augmentation techniques, including conventional noise augmentation using MUSAN and RIRs datasets and a unique speaker embedding-level mixup strategy for better optimization.
\end{abstract}


\section{Introduction}
\IEEEPARstart{A}{utomatic} speaker verification (ASV), which verifies and compares the speaker identity of an input utterance with enrollment data, is an extensively used biometric method. Depending on whether there are restrictions on the content of the speech, ASV systems can be divided into text-dependent ones, where the content of the spoken utterance is specified, or text-independent ones in which there are no restrictions on the linguistic content. With the development of deep learning, deep neural networks (DNN) have been successfully applied to both text-independent and text-dependent ASV tasks. The mainstream ASV models have evolved from statistical generative models such as Gaussian mixture models (GMM), universal background models (UBM)\cite{gmmubm} and i-vectors \cite{i-vector} to DNN-based discriminative models such as d-vector \cite{dvector,google-ge2e}, x-vector \cite{xvector1,xvector2,xvector-asp}, and r-vector \cite{r-vector}. One of the earliest DNN-based approaches is the d-vector \cite{dvector}, which takes frame-level acoustic features and their context as the input to predict the frame-level speaker-related representation, requiring post-processing to obtain the prediction of one utterance. More recently, DNN-based ASV systems utilize more efficient methods that directly extract utterance or segment-level speaker representative vectors (also called speaker embeddings) from the hidden layer of a DNN.

State-of-the-art DNN-based ASV systems can be broadly categorized into two groups: stage-wise approaches and fully end-to-end (E2E) approaches. The stage-wise approaches consist of a neural speaker encoder for extracting speaker embeddings as front-end processing and a back-end model for measuring the similarity of speaker embeddings, in which the processes in the two stages are trained separately. The front-end speaker encoder mainly consists of a frame-level feature extractor to capture discriminative speaker information, a pooling layer to accumulate the frame-level features into an utterance-level representation, and a loss function for the optimization.

Two typical stage-wise methods are time delay neural network (TDNN)-based x-vector and residual neural network (ResNet)-based r-vector. The x-vector utilizes TDNN as a frame-level feature extractor, followed by a statistics pooling layer to aggregate frame-level features into an utterance-level vector, which is then passed on to the following fully connected classifier. Instead of using TDNN, a 1-dimensional convolutional neural network, ResNet-based r-vector leverages 2-dimensional convolutional neural networks with residual connections, which is beneficial for building deeper neural networks as the frame-level feature extractor. Analogous to x-vector, r-vector also uses a temporal pooling layer for aggregating frame-level information into a single utterance-level speaker vector, and the following classifier uses the aggregated vector as input to categorize speakers. In addition to the frameworks of speaker encoders mentioned earlier, various approaches primarily related to the pooling methods and loss functions have also been proposed to improve the performance further. For example, a self-attention mechanism \cite{xvector-asp, vector-attention} was introduced into the pooling layer to calculate attentive statistics for more discriminative speaker representations. As for the loss functions of the speaker encoder, both metric learning-based loss functions such as triplet loss \cite{triplet-loss1,triplet-loss2} and margin-based softmax loss functions such as angular softmax loss \cite{angular-face,angular-spk} and additive margin softmax loss \cite{amsoftmax-face,lmcl} have shown superior performance in ASV tasks. It should be noted that the loss functions used in the speaker encoder are not verification-based ones but classification-based ones.

A back-end model, which measures the similarity of enrollment utterances and a testing utterance, is another crucial part of a modern ASV system. Cosine similarity is a simple but widely used back-end model for the front-end neural speaker encoder with a margin-based softmax loss function. Compared with cosine similarity, PLDA \cite{plda} is a more robust back-end model that can further disentangle speaker and channel factors in a latent space. Neural networks have also been used as the back-end model \cite{nplda}, and it was reported through a challenge competition \cite{voice-challenge} that this model outperforms the conventional PLDA model \cite{voice-challenge}. Moreover, there have been attempts to 1) combine a neural speaker encoder with a neural back-end, 2) design a fully E2E architecture that takes a pair of utterances as the input and generates the similarity score of trials directly from the inputs, and 3) optimize both of them by means of verification-based loss functions \cite{nplda, voice-challenge}.

Even though many novel ASV approaches are proposed every year, there is one scenario that has not been fully explored: a module that handles multiple enrollment utterances. For practical cases where it is more reliable to use multiple enrollment utterances, the operation of simply concatenating waveforms or averaging speaker embeddings is generally applied to these multiple enrollment utterances to obtain a single speaker representative vector \cite{cnceleb1,cnceleb2,reddots}, which is then treated as a single enrollment utterance. 

\begin{figure*}[!t]
\centering
\includegraphics[width=5in]{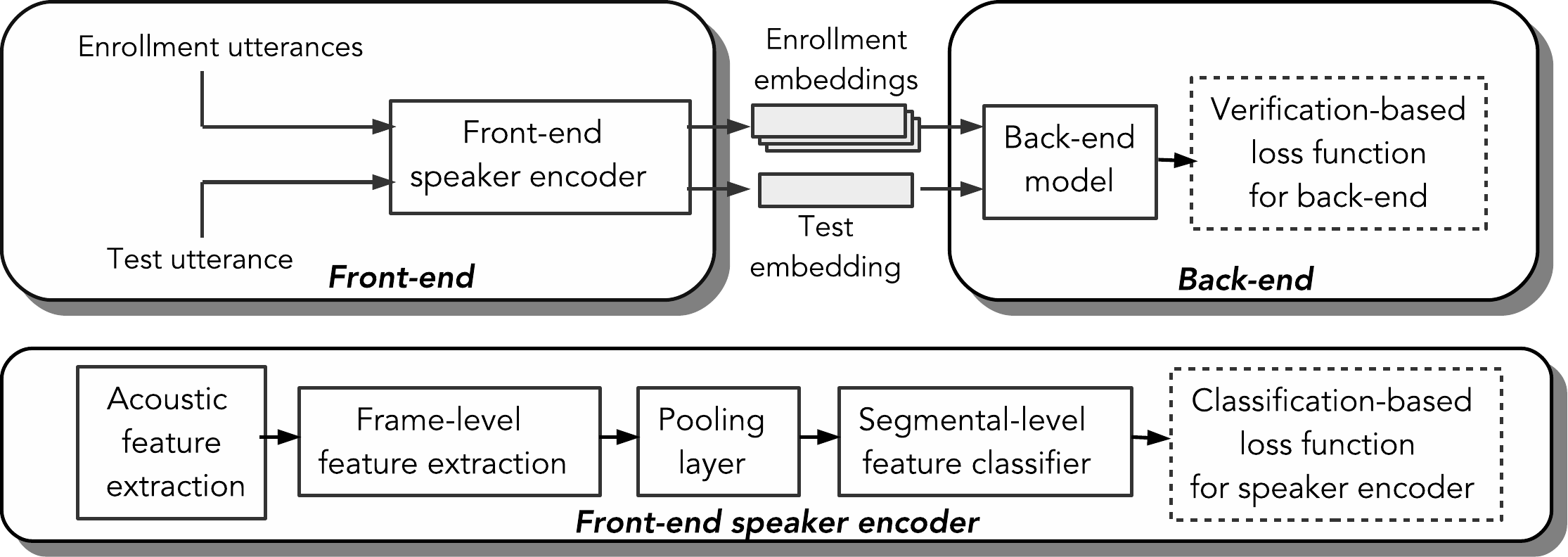}
\caption{General architecture of ASV.}
\label{fig:architecture-asv}
\vspace{-4mm}
\end{figure*}

In our previous study \cite{attention-backend}, we demonstrated that widely used simple operations such as concatenation and averaging for multiple enrollment utterances cannot efficiently leverage the intra-relationships among them. In order to better handle multiple enrollment utterances, a new neural back-end called attention back-end, which utilizes a scaled-dot self-attention mechanism \cite{attention-all-you-need} to learn the acoustic variations among a varying number of enrollment utterances, has been proposed, along with a feed-forward self-attention mechanism \cite{feed-forward-attention} to aggregate multiple speaker embeddings into a single speaker representative vector by using an adaptive weighted sum.

In this paper, to handle the case of multiple enrollment utterances more appropriately, we extend our earlier proposed attention back-end to an E2E approach by jointly fine-tuning the front-end speaker encoder and the attention back-end simultaneously.  We argue that the intra-speaker acoustic variations contained in multiple enrollment utterances can also overcome some of the intrinsic biases of the pretrained speaker encoder introduced by classification-based learning \cite{angular-face,aamsoftmax-face,amsoftmax-face} or metric learning \cite{contrastive-loss, triplet-loss1, triplet-loss2}. Additionally, considering the problem of heavily unbalanced positive and negative trials existing in the trials-sampling method \cite{attention-backend}, focal loss \cite{focal-loss} is selected to replace the binary cross-entropy loss to dynamically resample positive and negative trials and emphasize the contribution of difficult trials at the same time. We also introduce a unique embedding-level mixup \cite{mixup,mixup-spk}-based data augmentation strategy to increase the diversity of the training data.

The rest of this paper is presented as follows: In Section \ref{section:review}, we provide an overview of representative works, including the details of front-end speaker encoders, frequently used loss functions, and both conventional and neural back-end models. Section \ref{section:proposed} introduces the new fully E2E approach extending our previously proposed attention back-end to handle the case of multiple enrollment utterances. The method of sampling training trials for multiple enrollment utterances is also described in detail in this section. Section \ref{section:loss} lists the loss functions utilized in both the attention back-end and the proposed fully E2E model. Section \ref{section:augmentation} describes the signal-level and embedding-level data augmentation techniques we applied in ASV. In Section \ref{section:exp}, we report the experimental results of the proposed fully E2E method on the CNCeleb dataset for the case of multiple enrollments by comparing it with models that use separately-trained speaker encoder and back-end models as well as other SOTA E2E methods. We present an experiment in which we stack our proposed points gradually to determine the individual improvements. A deeper analysis of the effect of the number of enrollment utterances is also provided in this section. We conclude Section \ref{section:conclusion} with a brief summary.

\section{Related work}
\label{section:review}
\subsection{Stage-Wise Architecture of ASV}
An automatic speaker verification system usually consists of a front-end speaker encoder for extracting a speaker embedding and a back-end scoring model for giving a pair-wise score, which measures the similarity between enrollment utterances and the testing utterance. The whole system is illustrated at the top of Fig. \ref{fig:architecture-asv}.\footnote{In addition to the modules mentioned above, there is a preprocessing module that conducts centering, length normalization, and other operations on the speaker embedding.}

\subsubsection{Front-End Speaker Encoder}
\paragraph{Architecture of Speaker Encoder}
The speaker encoder, which projects a duration-varied utterance or acoustic features to a fixed-length discriminative speaker embedding, is an essential component for speaker verification systems. Generally speaking, the architecture of the speaker encoder (illustrated at the bottom of Fig. \ref{fig:architecture-asv}) can be divided into three parts, including a frame-level feature extractor, a pooling layer for aggregation, and a fully connected layers-based classifier. 

Many components have been proposed to extract discriminative speaker embeddings based on the above architecture. TDNN \cite{tdnn} and ResNet \cite{resnet} were the first to be applied as the backbone of speaker encoders in \cite{xvector1,xvector2,triplet-loss1}, and many variants \cite{xvector-asp,ecapa-tdnn,resnet34-fast,resnet34-performance} have appeared since then. This paper utilizes some representative speaker encoders in both the baseline systems and the proposed fully E2E approach (described in detail in Section \ref{section:baseline}).

\paragraph{Attention Mechanism for Speaker Encoder}
The attention mechanism is an essential component of the speaker encoder and can be incorporated into the frame-level feature extractor and the pooling layer for aggregation. Although there are many attention-based methods for speaker encoders, such as vector-based attentive pooling \cite{vector-attention}, multi-head attentive pooling \cite{self-multihead-attention}, multi-resolution multi-head attentive pooling \cite{mm-attention}, and others, in this section we review only the Squeeze-and-Excitation (SE) channel attention \cite{se} for frame-level feature extraction and the attentive statistics pooling layer \cite{xvector-asp} for aggregation, as they are used in most of the models in this paper. 

SE channel attention for frame-level feature extraction is introduced to integrate global information by re-scaling the outputs of convolutional layers that only focus on local information. In this paper we only examine the incorporation of the SE module with a 1-dimensional convolutional layer, but a 2-dimensional convolutional layer can also be used.

The squeeze operation first averages the frame-level feature map obtained from a 1-dimensional convolutional layer with $C$ channels denoted as
$\boldsymbol{M} = [\boldsymbol{m}_1, \cdots, \boldsymbol{m}_T] = [\boldsymbol{f}_1,\cdots,\boldsymbol{f}_C]\top \in \mathbb{R}^{C\times{T}}$ along the time domain, and obtains a vector $\boldsymbol{z}=\frac{1}{T}\sum_i^T\boldsymbol{m}_i$. Then, a compact excitation representation $\boldsymbol{s} = [s_1, \cdots, s_C]^\top = \sigma(\boldsymbol{W}_2^\top f(\boldsymbol{W}_1^\top\boldsymbol{z}+\boldsymbol{b}_1) + \boldsymbol{b}_2)$ is obtained by inputting $\boldsymbol{z}$ into two fully connected (FC) layers, where $\sigma (\cdot)$ is a sigmoid function and  $\boldsymbol{W}_1$ and $\boldsymbol{W}_2$ are $C\times{\frac{C}{r}}$ and ${\frac{C}{r}\times{C}}$ matrices that control the dimensions of hidden features with a ratio $r$. 

The final output of the SE module is obtained by rescaling the output at each channel $c$ with its channel attention weight $s_c$, that is, 
$\boldsymbol{\tilde{M}} = [\boldsymbol{\tilde{f}}_1,\cdots,\boldsymbol{\tilde{f}}_C]\top = [s_1\boldsymbol{f}_1,\cdots, s_C\boldsymbol{f}_C]\top$.

Attentive statistics pooling (ASP) applies attention mechanisms in the speaker verification task for aggregation. It concatenates the attention-based weighted mean and standard deviation of frame-level feature maps into a vector $\boldsymbol{c}=\text{Concat}(\boldsymbol{\tilde{\mu}},\boldsymbol{\tilde{\sigma}})$,
where $\boldsymbol{\tilde{\mu}}$ and $\boldsymbol{\tilde{\sigma}}$ are the weighted mean and standard deviation, respectively, and they are calculated with an attention weight $\alpha_t$ at each time step $t$ as follows:
\begin{align}
\label{formula:attention-weight}
    \boldsymbol{\tilde{\mu}} & = \sum_{t}^T\alpha_t\cdot \boldsymbol{m}_t, \\
    \boldsymbol{\tilde{\sigma}} & = \sqrt{\sum_{t}^T\alpha_t\boldsymbol{m}_t \odot 
    \boldsymbol{m}_t-\boldsymbol{\tilde{\mu}}\odot\boldsymbol{\tilde{\mu}}}, \\
    \alpha_t &= \frac{\text{exp}(\boldsymbol{v}^\top f(\boldsymbol{W}^\top\boldsymbol{m}_t+\boldsymbol{b})+k)}{\sum_t^T\text{exp}(\boldsymbol{v}^\top f(\boldsymbol{W}^\top\boldsymbol{m}_t+\boldsymbol{b})+k)},
\end{align}
where 
$\boldsymbol{v}$, $\boldsymbol{W}$, $\boldsymbol{b}$, and $k$ are learnable parameters and $f(\cdot)$ is a non-linear function such as ReLU or Tanh. 

\paragraph{Loss Function for Speaker Encoder}
The loss functions for the speaker encoder can be divided into two categories: metric learning (such as triplet loss \cite{triplet-loss1}) and softmax-based variants (such as plain softmax \cite{xvector1} and additive margin softmax (AM-softmax) \cite{amsoftmax-face,lmcl}). Compared to the former, softmax-based variants possess comparable performance and a simpler training pipeline without using any sampling method for training trials, which is usually used in metric learning \cite{mining}. Below, we review both plain softmax and AM-softmax because both of them are used in our experiments.

Plain softmax loss is used for classification, as
\begin{align}
    \label{formula:softmax}
    \mathcal{L}_{\text{softmax}} & = \nonumber \\ 
     -\frac{1}{N} & \sum_{n=1}^N \log \frac{e^{||\boldsymbol{w}_i||\cdot||\boldsymbol{e}_n||\cos \theta_{ni}}}{e^{||\boldsymbol{w}_i||\cdot||\boldsymbol{e}_n||\cos \theta_{ni}}+\sum_{j \neq i}^M e^{||\boldsymbol{w}_j||\cdot||\boldsymbol{e}_n||\cos \theta_{nj}}},
\end{align}
where $\theta_{ni}$ is the angle between the speaker embedding $\boldsymbol{e}_n$ of the $n$-th training sample and the weight vector $\boldsymbol{w}_i$ of its target class $i$. Here, the decision boundary between two different classes is decided by the length of the weight vector and the corresponding angle between the weight vector and the speaker embedding. This property leads to poor discrimination when using superficial cosine similarity as a back-end model. Therefore, the speaker encoder trained with the plain softmax loss function usually extracts the speaker embedding from the penultimate hidden layer, as in \cite{xvector2, xvector-asp}, and a PLDA back-end processes these embeddings for further improvement. 

Considering the undesirable properties of plain softmax loss, many softmax-based variants have been proposed \cite{angular-face,lmcl,aamsoftmax-face}. AM-softmax is a variant that can enlarge the distance between decision boundaries among different classes, as
\begin{equation}
    \label{amsoftmax}
    \mathcal{L}_{\text{am}}=-\frac{1}{N}\sum_{n=1}^N \log \frac{e^{s(\cos \theta_{ni} - m)}}{e^{s(\cos \theta_{ni} - m)}+\sum_{j \neq i}^M e^{s\cos \theta_{nj}}},
\end{equation}
where almost all annotations are the same as in Eq.\ (\ref{formula:softmax}) except for the two hyper-parameters $s$ and $m$, which rescale the value of $\cos \theta_{ni}$ and control the decision boundary margin among different classes, respectively. 

\subsubsection{Back-end}
\paragraph{Cosine Similarity}
Cosine score can measure the similarity of two speaker embeddings distributed in high-dimensional space.  It can be formulated as
\begin{equation}
\label{cosine similarity}
    \text{Cos}(\boldsymbol{e}_i,\boldsymbol{e}_j)=\frac{\boldsymbol{e}_i\top\boldsymbol{e}_j}{||\boldsymbol{e}_i||\cdot||\boldsymbol{e}_j||},
\end{equation}
where $\boldsymbol{e}_i$ and $\boldsymbol{e}_j$ are speaker embeddings from the same or different speakers. It has been frequently used as a scoring method for ASV systems trained using discriminative loss functions \cite{triplet-loss1,angular-spk,ecapa-tdnn,resnet34-fast}. In the case of multiple enrollment utterances, a frequently used operation is to average the speaker embeddings $\boldsymbol{c}_n = \frac{1}{L}\sum_{l=1}^L \boldsymbol{e}_l$, where $L$ is the number of enrollment utterances per speaker. Alternatively, a speaker embedding may be extracted from concatenated waveforms, or from acoustic features \cite{attention-backend}.

\paragraph{Probabilistic Linear Discriminant Analysis}
Probabilistic Linear Discriminant Analysis (PLDA) is another popular method for scoring in speaker verification---not only in statistics-based methods such as i-vector but also in neural network-based methods such as x-vector. There are many variants of PLDA models, but here we only review the simple Gaussian PLDA (GPLDA) model:

\begin{equation}
\label{PLDA}
    \boldsymbol{e}_{n}=\boldsymbol{e}_\mu+\boldsymbol{F}\boldsymbol{\omega}_n+\boldsymbol{\epsilon}_{n},
\end{equation}
where $\boldsymbol{e}_\mu$ represents the center vector of the training speaker embeddings. In GPLDA, both the latent variable $\boldsymbol{\omega}_n$ and residual term $\boldsymbol{\epsilon}_n$ for the $n$-th utterance are assumed to follow Gaussian distributions, that is, $\boldsymbol{\omega}_n \sim \mathcal{N}(0, \mathbf{I})$ and $\boldsymbol{\epsilon}_n \sim \mathcal{N}(0, \mathbf{\Sigma})$, where $\mathbf{I}$ and $\boldsymbol{\Sigma}$ represent identity and covariance matrices, respectively. The score of a pair of speaker embeddings is given as follows for a single-session enrollment case:
\begin{align}
    \label{plda-scoring}
    P(\boldsymbol{e}_{i},\boldsymbol{e}_{j}) & = \boldsymbol{e}_{i}^\top \boldsymbol{Q}\boldsymbol{e}_{i}+\boldsymbol{e}_{j}^\top \boldsymbol{Q}\boldsymbol{e}_{j}+2\boldsymbol{e}_{i}^\top \boldsymbol{P}\boldsymbol{e}_{j}, \\
    \boldsymbol{P} & = \boldsymbol{\Sigma}^{-1}_{tot}\boldsymbol{\Sigma}_{ac}(\boldsymbol{\Sigma}_{tot}-\boldsymbol{\Sigma}_{ac} \boldsymbol{\Sigma}^{-1}_{tot}\boldsymbol{\Sigma}_{ac})^{-1},  \\ 
    \boldsymbol{Q} & =\boldsymbol{\Sigma}^{-1}_{tot}-(\boldsymbol{\Sigma}_{tot}-\boldsymbol{\Sigma}_{ac}\boldsymbol{\Sigma}^{-1}_{tot}\boldsymbol{\Sigma}_{ac})^{-1}, \\
    \boldsymbol{\Sigma}_{tot} & =\boldsymbol{F}\boldsymbol{F}^\top+\boldsymbol{\Sigma}, \\ 
    \boldsymbol{\Sigma}_{ac} & =\boldsymbol{F}\boldsymbol{F}^\top,
\end{align}
where $\{\boldsymbol{\mu},\boldsymbol{F},\boldsymbol{\Sigma}\}$ are trainable parameters that are estimated using of the Expectation-Maximization (EM) algorithm. 

There are several ways of handling the multi-session enrollment case with PLDA \cite{kong-plda, ville-plda}. In light of the results reported in \cite{kong-plda,ville-plda}, we adopt the average of multiple speaker embeddings in our baseline models due to its better performance.

\paragraph{Neural Probabilistic Linear Discriminant Analysis}
Inspired by the PLDA pair-wise scoring method, \cite{nplda} proposed a neural network model to simulate the behavior of PLDA, called neural PLDA (NPLDA). Referring to the PLDA implementation in Kaldi \cite{kaldi}, NPLDA uses two fully connected layers as substitutes for the LDA and centering processing. Then, two semi-orthogonal matrices are leveraged to represent the corresponding $\boldsymbol{Q}$ and $\boldsymbol{P}$ in Eq.\ (\ref{plda-scoring}), respectively, which measure the similarity between the enrollment and testing speaker embeddings. According to \cite{nplda}, initializing the parameters of NPLDA from a well-trained PLDA model is an important trick to obtain a large improvement over the conventional PLDA model. For the scenario of multiple enrollment utterances, there is no special description in \cite{nplda}. Therefore, we average all enrollment speaker embeddings as one speaker-representative vector and use NPLDA to compute the score of this vector and the testing speaker embedding as a baseline system.

\paragraph{Loss Function for Neural Back-end}
In general, a back-end model in speaker verification measures the similarity of a trial pair, which contains one or more enrollment utterances and one testing utterance. Therefore, the following binary cross-entropy (BCE) loss function is a good choice to evaluate how well the back-end model is trained:
\begin{align}
    \mathcal{L}_{\text{BCE}} = & -\sum_{k=1}^{K}[\mathds{1}_{y_k=1}\mathrm{log}\sigma(s_k)+ 
    \mathds{1}_{y_k\ne1}\mathrm{log}(1-\sigma(s_k))],
\end{align}
where $s_k$ is the similarity score of the $k$-th trial pair, and $\mathds{1}$ is an indicator function. The ground-truth label $y_k=1$ denotes that the trial pair is positive, i.e., that both the enrollment and test trial are from the same speaker, while $y_k \neq 1$ denotes that it is negative. 

The normalized minimum detection cost function (minDCF) formulated by (\ref{formula:norm-mindcf}), is frequently used to evaluate the performance of ASV systems by considering practical cases, as
\begin{align}
\label{formula:norm-mindcf}
    C_{norm}(\beta,\eta)=P_{miss}(\eta)+\beta P_{fa}(\eta),
\end{align}
where $\eta$ is a threshold value to make a hard decision. $P_{miss}$ and $P_{fa}$ are the probability of missing and false alarms decided by $\eta$, respectively. $\beta$ is a predefined application-dependent number represented by:
\begin{align}
\label{formula:beta}
    \beta = \frac{C_{fa}(1-P_{target})}{C_{miss}P_{target}},
\end{align}
where $C_{miss}$ and $C_{fa}$ are the costs assigned to miss and false alarms, respectively, and $P_{target}$ is the prior probability of a target trial \cite{nplda-e2e}.

However, it cannot be used as a loss function due to its non-differentiable property. There has been an attempt \cite{nplda} to optimize the minDCF metric directly for training so that it can be introduced in gradient-based optimization by approximating it in a differentiable way, as:
\begin{align}
\label{formula:a-mindcf}
    \mathcal{L}_{\text{aDCF}}=\min\limits_{\eta}C_{norm}^{(soft)}(\beta, \eta),
\end{align}
where the parameter $\theta$ is differentiable by introducing it into an approximate version of $P_{miss}$ and $P_{fa}$, which is calculated as:
\begin{align}
    P_{miss}^{(soft)}(\eta) & = \frac{\sum_{k=1}^Ky_k[1-\sigma(\delta(s_k-\eta))]}{\sum_{k=1}^Ky_k}, \\
    P_{fa}^{(soft)}(\eta) & = \frac{\sum_{k=1}^K(1-y_k)\sigma(\delta(s_k-\eta))}{\sum_{k=1}^K(1-y_k)}.
\end{align}
Here, $k$ is the trial index, $s_k$ is the system score, $y_k$ denotes the ground-truth label for trial $k$, and $\sigma$ denotes a sigmoid function. $K$ is the total number of trials in the mini-batch over which the cost is computed. By choosing a large enough value for the warping factor $\delta$, the approximation can be made arbitrarily close to the actual detection cost function for a wide range of thresholds \cite{nplda}.

\begin{figure}[!t]
\centering
\includegraphics[width=3.4in]{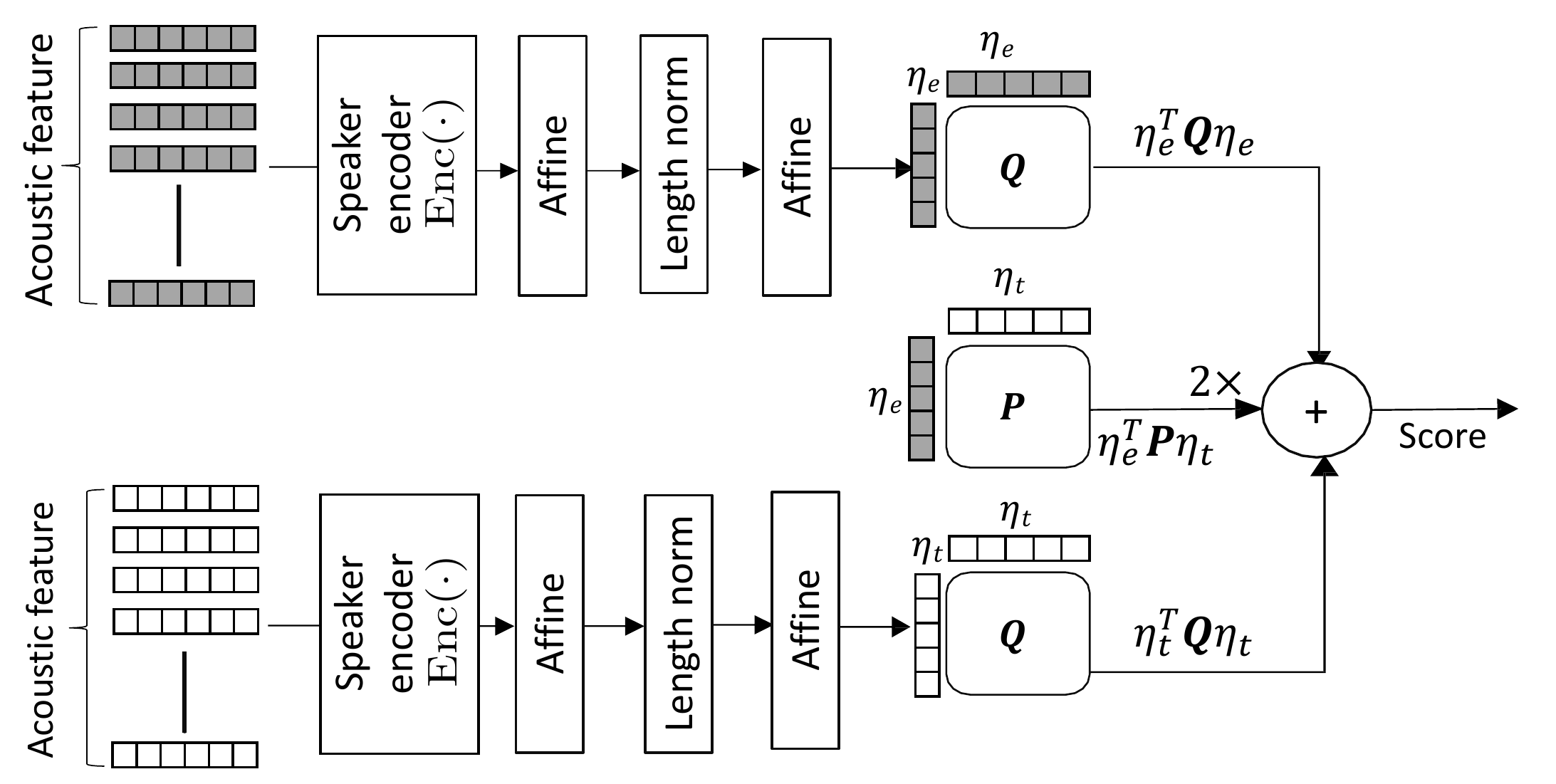}
\caption{NPLDA-E2E architecture. The speaker encoder and the NPLDA are trained independently at first, and the two models are then concatenated for fine-tuning in E2E style.}
\label{fig:architecture-nplda-e2e}
\vspace{-4mm}
\end{figure}

\subsection{Fully End-to-End Method}

There are several definitions of E2E in the context of speaker verification. For example, some papers \cite{triplet-loss1,end2end-triplet,end2end-angular} have described their systems, which train a speaker encoder with a discriminative loss function and use cosine score to measure the similarity between enrollment and testing speaker embeddings extracted from the well-trained speaker encoder, as E2E speaker verification systems. However, in the current work, \textit{we refer to a system that is fully trained directly from waveforms or acoustic features and optimized with a speaker verification metric, directly outputting the similarity score for a trial pair in the testing stage, as a fully E2E system (e.g., generalized E2E (GE2E) \cite{google-ge2e}, E2E NPLDA \cite{nplda-e2e}, and so on \cite{end2end-dnn,end2end-complete,google-e2e}}). In this section, we review only the recently proposed representative NPLDA-E2E architecture, which demonstrated a successful application in the VOiCES challenge \cite{voice-challenge}.

This approach concatenates a speaker encoder, which is TDNN in \cite{nplda-e2e}, and an NPLDA back-end model in order to optimize an ASV system with the approximated DCF loss function in Eq.\ (\ref{formula:a-mindcf}) simultaneously. Figure \ref{fig:architecture-nplda-e2e} illustrates the enrollment and evaluation process of this fully E2E model. Additionally, a novel sampling method of trials is proposed in \cite{nplda-e2e} to simulate the enrollment and evaluation of the testing stage during the training stage, which improves the performance due to the resultant consistency. In the case of multiple enrollment utterances, the enrollment speaker embeddings are simply averaged before being processed by the NPLDA layers.

\begin{figure}[!t]
\centering
\includegraphics[width=3in]{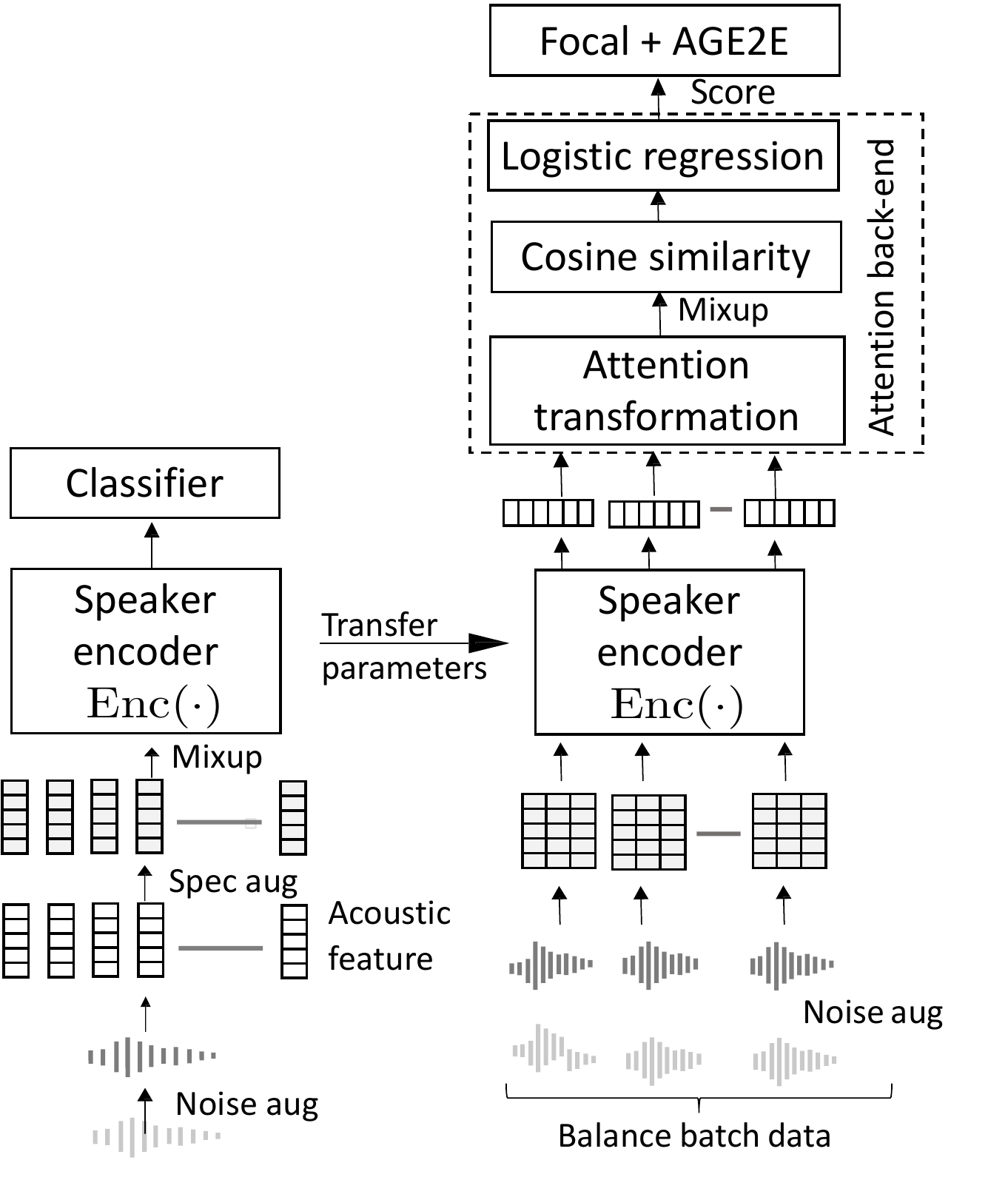}
\caption{Architecture of proposed fully E2E method. The speaker encoder is trained as a classification task with noise augmentation \cite{musan, rirs}, SpecAugment \cite{specaug}, and a mixup strategy \cite{mixup, mixup-spk} at first stage. The well-trained speaker encoder is then concatenated with a randomly initialized attention back-end for the second E2E fine-tuning stage.}
\label{fig:proposed-architecture}
\vspace{-4mm}
\end{figure}

\section{End-to-End Joint Model of Speaker Encoder and Attention Back-end}
\label{section:proposed}
As discussed in the previous section, a practical speaker verification system is divided into two parts: a front-end speaker encoder for extracting a speaker embedding and a back-end for scoring a pair-wise trial. However, the independent optimization of these separate components is prone to falling into an undesirable local minimum \cite{end2end-complete,nplda-e2e}. Several joint-optimization methods have been proposed to solve this problem by combining the front-end speaker encoder and the back-end scoring model \cite{google-e2e, google-ge2e, nplda-e2e}, and have shown substantial improvements compared to separate speaker encoders and back-ends. Nevertheless, none of the above-mentioned models have the capability to make full use of the intra-relationships among multiple enrollment utterances. In this section, we propose a new architecture for a fully E2E model that not only follows the E2E training pipeline but also leverages the intra-relationships among multiple enrollment utterances. Considering the difficulty of training a fully E2E neural network directly, we adopt a two-stage training strategy (shown in Fig. \ref{fig:proposed-architecture}) in which a speaker encoder is firstly optimized with a softmax or AM-softmax loss function and then the parameters of the speaker encoder are transferred to the fully E2E model for fine-tuning. 

\subsection{Discriminative Speaker Encoder}
In the first stage of optimization, many common speaker encoders (e.g. TDNN, ResNet) can be trained as a good seed model for the second fine-tuning stage. To improve the robustness against complicated scenarios such as noisy environments, we make use of a few data augmentation tricks, including noise augmentation, mixup, and SpecAug \cite{specaug}, to increase the diversity of training data.

\subsection{Fully End-to-End Fine-tuning}
\label{section:e2e-finetune}
In the fine-tuning stage, the well-trained speaker encoder is concatenated with a randomly initialized attention back-end component, and the concatenated model is optimized using an E2E training method with a verification-based loss function. The input of the model is the acoustic features of utterances, and the output is the scores of pair-wise trials. We implemented multi-level aggregation methods, including frame-level attention aggregation in some speaker encoder components and utterance-level attention aggregation in the attention back-end component, to leverage the intra-relationships of multiple enrollment utterances. 

\subsubsection{Trial Sampling Method}
\begin{figure}[!t]
\centering
\includegraphics[width=3.4in, trim={0 0 0 0},clip]{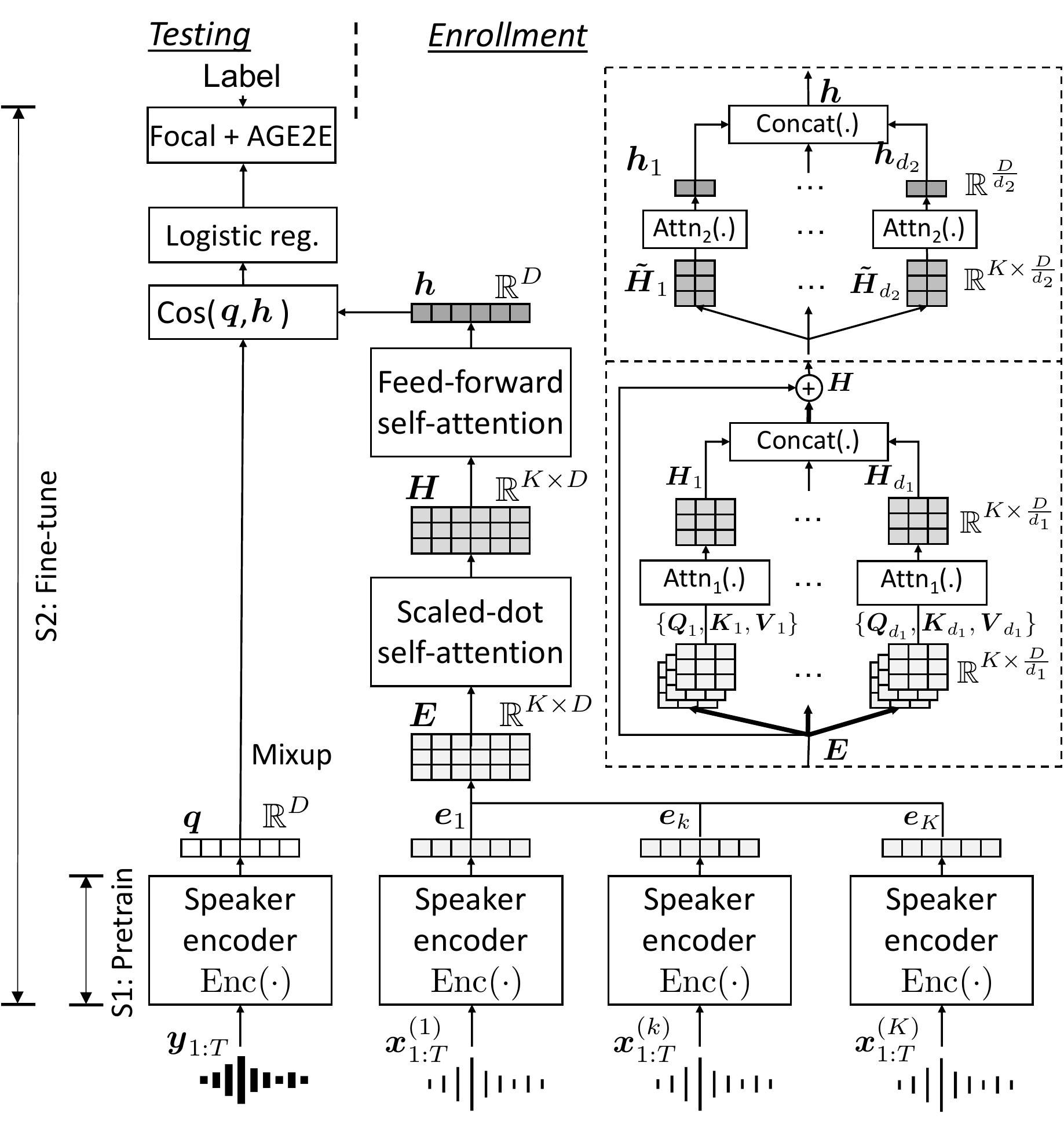}
\caption{Proposed method showing expanded detail of the attention back-end structure. The speaker encoder is pretrained at stage 1 (S1). The whole system, which consists of the pretrained speaker encoder and a randomly initialized attention back-end, is fine-tuned at stage 2 (S2).}
\label{fig:attention-backend}
\vspace{-4mm}
\end{figure}

When training the fully E2E model, we adopt a sampling method that allows each speaker in the same mini-batch to possess an equal number of utterances instead of randomly sampling utterances in the same manner as the first training stage. Speaker embeddings can be extracted from the acoustic features of the utterances in a mini-batch by means of a speaker encoder module (e.g. TDNN, ResNet) in our fully E2E method, as Fig. \ref{fig:attention-backend} shows. Considering that speakers in a mini-batch can be viewed as either target or non-target speakers, we rearrange the speaker embeddings in one mini-batch to define various positive and negative pairs. Table~\ref{fig:train} gives a simple example to illustrate our method of rearrangement. Assuming there are three speakers, $A$, $B$, and $C$, and four utterances whose indices range from 1 to 4 per speaker, we select one utterance as a test trial and use the others of the same speaker can be used as enrollments for positive pairs. Negative pairs can be formed by selecting a test trial denoted as (\PTrial) from one speaker and enrollments denoted as (\Nenroll, \Nenroll, \Nenroll) from other speakers.

\begin{table}[t]
\scriptsize
    \centering
    \caption{Composition of pairs of (test-trial, enrollment-data) for training back-end model and ground-truth labels from mini-batch. A, B, and C are speaker IDs, and 1, 2, 3, and 4 are his/her audio IDs. \PTrial and \Penroll denote test and enrollment audio files, respectively. \label{fig:train}}
    \vspace{0mm}
    \setlength{\tabcolsep}{3pt}
    {
    \begin{tabular}{ccccccccccccccccc}
 \cmidrule{2-13} 
 & \multicolumn{4}{c}{$A$} & \multicolumn{4}{c}{$B$} & \multicolumn{4}{c}{$C$} \\
 \cmidrule{2-13} 
 & 1 & 2 & 3 & 4 & 1 & 2 & 3 & 4 & 1 & 2 & 3 & 4 & & Test & Enroll & Label\\
 \cmidrule{2-13}
 \multirow{10}*{\rotatebox[origin=c]{90}{\textbf{Trials to be used for training}}} & \PTrial & \Penroll  & \Penroll & \Penroll & & & & & & & & & & $\boldsymbol{q}_{A1}$ & $\boldsymbol{h}_{A1}$ & P\\
 
 & \PTrial &  &  &  &  & \Nenroll  & \Nenroll & \Nenroll  & &  & & & & $\boldsymbol{q}_{A1}$ & $\boldsymbol{h}_{B1}$ & N \\
 
 & \PTrial &  &  &  &  & &  & & & \Nenroll  & \Nenroll & \Nenroll &  & $\boldsymbol{q}_{A1}$ & $\boldsymbol{h}_{C1}$ & N \\

 & \Penroll & \PTrial & \Penroll  & \Penroll & & & & & & & & & & $\boldsymbol{q}_{A2}$ & $\boldsymbol{h}_{A2}$ & P\\
 
 & & \PTrial &  &  & \Nenroll & & \Nenroll  & \Nenroll &  &  & & & & $\boldsymbol{q}_{A2}$ & $\boldsymbol{h}_{B2}$ & N \\
 
 & & \PTrial &  &  &  &  &  & & \Nenroll & & \Nenroll & \Nenroll & &  $\boldsymbol{q}_{A2}$ & $\boldsymbol{h}_{C2}$ & N \\
 
 & & & & &  & & \vdots & & & & &  & & \\

 & \Nenroll & \Nenroll & \Nenroll & &  &  &  &  &  &  & & \PTrial  & & $\boldsymbol{q}_{C4}$ & $\boldsymbol{h}_{A4}$ & N \\
 
 & & & &  & \Nenroll  & \Nenroll & \Nenroll &   &  &  &  & \PTrial & & $\boldsymbol{q}_{C4}$ & $\boldsymbol{h}_{B4}$ & N \\
 
 & &  &  &   & & & & &  \Penroll & \Penroll & \Penroll & \PTrial & & $\boldsymbol{q}_{C4}$ & $\boldsymbol{h}_{C4}$ & P\\
 \cmidrule{2-13}
\end{tabular}
}
\vspace{-4mm}
\end{table}

\subsubsection{Utterance-level Attention}
After sampling and rearranging numerous trials from the data in a batch, the training process is the same as the inference process. Therefore, in the rest of this section, we only describe the details of how inference is implemented in the fully E2E model. Assuming a speaker has $K$ enrollment utterances, which may vary for different speakers, $K$ speaker embeddings $\{\boldsymbol{e}_1, \cdots, \boldsymbol{e}_K\}$ transformed by speaker encoder $\text{Enc}(\cdot)$ are stacked to form a speaker embedding matrix $\boldsymbol{E}\in\mathbb{R}^{K\times{D}}$ as input to the following module. Here, $D$ is the dimension of the speaker embedding generated by the speaker encoder. Since we argue that the relationship among multiple enrollment utterances is important in order to make full use of them, we utilize multi-head scaled dot self-attention (SDSA) \cite{attention-all-you-need} here to project them to a latent matrix denoted by $\boldsymbol{H}\in\mathbb{R}^{K\times{D}}$, as
\begin{equation}
    \label{equa:multihead}
    \boldsymbol{H} = {\rm{Concat}}(\boldsymbol{H}_1, \boldsymbol{H}_2, \cdots , \boldsymbol{H}_{d_1}) \boldsymbol{W}^O + \boldsymbol{E},
\end{equation}
where $d_1$ represents the number of heads in SDSA. The $i$-th sub-matrix $\boldsymbol{H}_i\in\mathbb{R}^{K\times\frac{D}{d_1}}$ is computed by
\begin{equation}
    \label{equa:attention_transform}
    \boldsymbol{H}_i = \text{Attn}_1(\boldsymbol{Q}_i, \boldsymbol{K}_i, \boldsymbol{V}_i) = \text{Softmax}(\frac{\boldsymbol{Q}_i\boldsymbol{K}_i^\top}{\sqrt{{D}/{d_1}}})\boldsymbol{V}_i.
\end{equation}
As described in \cite{attention-all-you-need}, $\boldsymbol{Q}_i$, $\boldsymbol{K}_i$, and $\boldsymbol{V}_i$ are query, key, and value matrices, respectively, which can be obtained by applying linear transformations with trainable parameter matrices of $\boldsymbol{W}_i^Q$, $\boldsymbol{W}_i^K$, and $\boldsymbol{W}_i^V$ on input matrix $\boldsymbol{E}$. Through the multi-head mechanism in Eq. (\ref{equa:multihead}), the information of different positions of $\boldsymbol{E}$ can be attended from different representation sub-spaces \cite{attention-all-you-need}.

As $K$ is a varying number, it is necessary to aggregate the latent matrix $\boldsymbol{H}$ to a speaker representative vector $\boldsymbol{h}$. In order to do this, another attention mechanism called multi-head feed-forward self-attention (FFSA) \cite{feed-forward-attention} is utilized, as
\begin{equation}
    \label{equa:multihead_aggregation}
    \boldsymbol{h} = {\text{Concat}}(\boldsymbol{h}_1, \boldsymbol{h}_1, \cdots, \boldsymbol{h}_{d_2}),
\end{equation}
where $d_2$ represents the number of heads in FFSA. For each head vector $\boldsymbol{h}_j\in\mathbb{R}^{\frac{D}{d_2}}$, it is derived as
\begin{equation}
   \label{equa:multihead_aggregation_2}
    \boldsymbol{h}_j = \text{Attn}_2{(\boldsymbol{\tilde{H}}_j)}=\text{Softmax}(\boldsymbol{v}_j^\top \text{Tanh}(\boldsymbol{W}_j \boldsymbol{\tilde{H}}_j^\top)) \boldsymbol{\tilde{H}}_j,
\end{equation}
where the adaptive weights of $\boldsymbol{h}_j$ are learned by a small feed forward neural network. The trainable parameters $\boldsymbol{W}_j\in\mathbb{R}^{D_{2}\times{\frac{D}{d_2}}}$ and $\boldsymbol{v}_j\in\mathbb{R}^{D_{2}}$ are applied on input $\boldsymbol{\tilde{H}}_{j}$ in order, which is a sub-matrix of $\boldsymbol{H}$, with successive non-linear transformation $\text{Tanh}$ and $\text{Softmax}$, respectively.

\subsubsection{Score Calibration}
After aggregating the enrollment embeddings via attention mechanisms, the similarity between an evaluating speaker embedding $\boldsymbol{q}$, which is extracted by the speaker encoder, and the claimed speaker representative vector $\boldsymbol{h}$ is measured by cosine score, as
\begin{align}
\label{equa:cosine-score}
    \mathrm{Cos}(\boldsymbol{q},\boldsymbol{h}) &=\frac{\boldsymbol{q} ^\top \boldsymbol{h}}{||\boldsymbol{q}||\cdot||\boldsymbol{h}||}.
\end{align}
Next, a logistic regression (LR) is used for better calibration, as
\begin{align}
\label{equa:lr}
    P(\boldsymbol{q},\boldsymbol{h}) 
    &= \frac{1}{1 + \exp^{-s}} 
    = \frac{1}{1 + \exp^{-a \, \text{Cos}(\boldsymbol{q},\boldsymbol{h}) - b}},
\end{align}
where $a$ and $b$ are learnable parameters of LR and $P(\boldsymbol{q},\boldsymbol{h})$ denotes the probability that the speaker of the evaluating utterance is the same as that of the enrollment utterances.

\section{Loss Functions}
\label{section:loss}
Using a classification-based loss function in the speaker encoder training stage would be inconsistent with the verification process. Therefore, in the fine-tuning stage, we utilize the weighted sum of the attention mechanism-based generalized E2E (AGE2E) loss and the binary focal loss as the criterion to supervise the fine-tuning of the proposed method, as
\begin{equation}
\label{formula:loss}
\mathcal{L}=\lambda \mathcal{L}_{\mathrm{AGE2E}} + (1-\lambda) \mathcal{L}_{\mathrm{FOCAL}},
\end{equation}
where AGE2E loss is a variant of the generalized E2E (GE2E) loss and is designed to better leverage the training trial pairs selected by our trial sampling method in each mini-batch \cite{attention-backend}. At the same time, the focal loss is expected to alleviate the issue of unbalanced positive and negative training pairs stemming from the trial sampling method. It also encourages the model to focus on difficult trial pairs. The hyper-parameter $\lambda$ is set to 0.6 in this paper.

\subsection{Attention-based Generalized End-to-End Loss}
The AGE2E loss is used in our model to reduce within-class variance and enlarge between-class distances. Unlike BCE, AGE2E loss exploits the information of the trial pairs in the mini-batch. Let us consider the example in Table~\ref{fig:train} and define $\boldsymbol{q}_{lm}$ as a speaker embedding vector extracted from the $m$-th test trial of speaker $l$. Let us also define $\boldsymbol{h}_{nm}$ as the vector $\boldsymbol{h}$ of Eq.\ (\ref{equa:multihead_aggregation}) extracted from the enrollment set that contains multiple audio files uttered by speaker $n$ (except the $m$-th test trial).  
Given all the $\boldsymbol{q}_{lm}$ and $\boldsymbol{h}_{nm}$ from the mini-batch, the AGE2E loss is computed by
 \begin{align}
 \label{ge2e}
 \mathcal{L}_{\mathrm{AGE2E}} & = - \sum_{^{\forall} l, m} \log \frac{\exp^{P ( \boldsymbol{q}_{lm},\boldsymbol{h}_{lm} )}}{\sum_{^{\forall} n} \exp^{P ( \boldsymbol{q}_{lm},\boldsymbol{h}_{nm} )}}.
 \end{align}
Although the definition above is similar to the GE2E loss \cite{google-ge2e}, a crucial difference is on $\boldsymbol{h}$. While the original GE2E simply averages multiple embeddings per speaker and uses the resulting centroid as $\boldsymbol{h}$, we utilize the attention mechanism in Eqs.\ (\ref{equa:multihead_aggregation}\textendash\ref{equa:multihead_aggregation_2}) to compute the $\boldsymbol{h}$. This is why we refer to it as attention-based GE2E.  The attention-based $\boldsymbol{h}$ is arguably more powerful than the simple centroid to represent the speaker in the embedding space.

\subsection{Focal Loss}
As discussed in \cite{attention-backend}, the trial sampling strategy results in a heavily unbalanced data ratio of positive and negative trials. Furthermore, BCE loss in \cite{attention-backend} treats difficult and easy trials equally, which is not ideal in our use case. We use focal loss \cite{focal-loss, focal-loss-spk} as a substitute for BCE loss to alleviate this problem. 

The focal loss used for a binary classification task is defined on the basis of the BCE loss and can be written as
\begin{align}
\label{formula:focal-loss}
    &\mathcal{L}_{\text{FOCAL}} \nonumber \\
    =& - \sum_{^{\forall} l, m, n} [\mathds{1}_{ l=n }\alpha (1 - P(\boldsymbol{q}_{lm},\boldsymbol{h}_{nm}))^\gamma \log P(\boldsymbol{q}_{lm},\boldsymbol{h}_{nm}) \nonumber \\
    & + \mathds{1}_{ l \neq n }(1-\alpha)P(\boldsymbol{q}_{lm},\boldsymbol{h}_{nm})^\gamma \log (1-P(\boldsymbol{q}_{lm}, \boldsymbol{h}_{nm}))],
\end{align}
where the hyper-parameter $\alpha$ controls the balance between positive and negative classes, and  $(1-P(\boldsymbol{q}_{lm},\boldsymbol{h}_{nm}))^\gamma$ and $P(\boldsymbol{q}_{lm},\boldsymbol{h}_{nm})^\gamma$ are modulating factors that dynamically adjust the weight of the loss for each trial pair. The \emph{focusing} hyper-parameter $\gamma \geq 0$ further controls the scale of the modulating factors. When $\gamma=0$ and $\alpha=0.5$, the focal loss function is equivalent to the BCE loss.

The focal loss encourages the model to focus on difficult pairs. Let us consider the case where $l\neq n$, i.e., the test trial and enrollment utterances, are from different speakers. If the input pair is easy to classify, the E2E model should produce a small $P(\boldsymbol{q}_{lm},\boldsymbol{h}_{nm})$, and the resulting modulating factor  $P(\boldsymbol{q}_{lm},\boldsymbol{h}_{nm})^\gamma$ also becomes small. Consequently, the loss from this easy pair is down-weighted. In contrast, if the model produces a large $P(\boldsymbol{q}_{lm},\boldsymbol{h}_{nm})$ close to 1, the modulation factor $P(\boldsymbol{q}_{lm},\boldsymbol{h}_{nm})^\gamma$ will also be close to 1, and the loss from this difficult pair is not down-weighted. Thus, the difficult trial pair contributes more to the training loss as well as to the gradient. 
By replacing the BCE loss with focal loss and adjusting the two hyper-parameters $\{\alpha, \gamma\}$, we observed that this mechanism could indeed help the model converge to a better state in the experiments.

\section{Multi-level Data Augmentation}
\label{section:augmentation}
Data augmentation plays a key role in building a state-of-the-art ASV system \cite{xvector1,resnet34-fast,mixup-spk,ecapa-tdnn}. In addition to simple noise augmentation, SpecAug \cite{specaug} and speaker augmentation \cite{spk-aug} are frequently used to increase the robustness of an ASV system. Moreover, a mixup strategy, which minimizes the vicinal risk \cite{mixup} in training, has also been investigated at both the signal and acoustic feature levels \cite{mixup-spk}. Instead of the implementation in \cite{mixup-spk}, we applied the mixup strategy at the speaker embedding level to augment the training data in the fine-tuning stage. As far as we know, this is the first time an embedding-level mixup strategy has been applied to the ASV task. We explain it in detail after first introducing the standard signal-level augmentation method.

\subsection{Signal-level Augmentation}
In our proposed method, we use the MUSAN \cite{musan} and RIRs \cite{rirs} corpora for signal-level data augmentation. Here, we follow the KALDI \cite{kaldi} CNCeleb recipe to implement data augmentation in both the speaker encoder pretraining stage and the fully E2E fine-tuning stage.

We also use mixup \cite{mixup, mixup-spk} from acoustic features is also used in the pretraining stage to improve the generalization performance of the model. Mixup can be viewed as a data augmentation method, and we use it here to create new vicinal samples from the acoustic feature level. The mixup operation can be written as
\begin{align}
    \tilde{\boldsymbol{x}}=\beta_1 \boldsymbol{x}_1 + (1-\beta_1) \boldsymbol{x}_2, \\
    \tilde{\boldsymbol{y}}=\beta_1 \boldsymbol{y}_1 + (1-\beta_1) \boldsymbol{y}_2,
\end{align}
where $\{\boldsymbol{x}_1, \boldsymbol{y}_1\}$ and $\{\boldsymbol{x}_2, \boldsymbol{y}_2\}$ are the pairs of acoustic features and one-hot encodings of target labels from two randomly selected data samples. The weight $\beta_1$ can be generated from a $\text{Beta}$ distribution with a hyper-parameter $\psi_1$, i.e., $\beta_1 \sim \text{Beta}(\psi_1, \psi_1)$. In all experiments in this paper, the parameter $\psi_1$ is set to 1.0.

The new data sample $\{\tilde{\boldsymbol{x}}, \tilde{\boldsymbol{y}}\}$ is used as augmented data to train the neural speaker encoder for extracting speaker embeddings. In the speaker encoder pretraining stage, AM-softmax or other margin-based variants on the augmented data can be written as
\begin{align}
    \mathcal{L}_{am}(\tilde{\boldsymbol{x}},\tilde{\boldsymbol{y}})=\beta_1 \mathcal{L}_{am}({\tilde{\boldsymbol{x}}},\boldsymbol{y}_1) + (1-\beta_1)\mathcal{L}_{am}({\tilde{\boldsymbol{x}}}, \boldsymbol{y}_2).
\end{align}

\subsection{Embedding-level Augmentation}
\label{sec:embeddingaugmentation}
Signal-level augmentation has exhibited success in the automatic speaker verification task \cite{xvector1, ecapa-tdnn, mixup-spk}. However, in our system, we hypothesize that the mixup augmentation at the embedding level is also helpful, since the distribution of training pairs in our fully E2E method is heavily unbalanced due to adopting the sampling method in \ref{section:e2e-finetune}. The embedding-level mixup strategy is applied on the speaker embedding $\boldsymbol{q}$ of the test trial, as shown in Fig. \ref{fig:attention-backend}. It is inserted before computing the cosine similarity between $\boldsymbol{h}$ and $\boldsymbol{q}$. This operation can be formulated as
\begin{align}
    \tilde{\boldsymbol{q}} = \beta_2 \boldsymbol{q}_1 + (1 - \beta_2) \boldsymbol{q}_2, \\
    \tilde{y} = \beta_2 y_1 + (1 - \beta_2) y_2,
\end{align}
where $\boldsymbol{q}_1$ and $\boldsymbol{q}_2$ are testing speaker embeddings of the test branch in Fig. \ref{fig:attention-backend}. $y_1$ and $y_2$ are the corresponding labels of the pairs of these testing speaker embeddings and a speaker representative vector $\boldsymbol{h}$. The weight $\beta_2$ is also generated from a Beta distribution with a hyper-parameter $\psi_2$, which is set to 1.0 in this paper. The similarity between the mixup augmented testing speaker embedding $\tilde{\boldsymbol{q}}$ and the speaker representative vector $\boldsymbol{h}$ is computed by using cosine score followed by logistic regression for score calibration. Therefore, the loss function for embedding-level mixup can be written as
\begin{align}
    \mathcal{L}(\tilde{\boldsymbol{q}},\boldsymbol{h},\tilde{y})=\beta_2 \mathcal{L}({\boldsymbol{\tilde{q}}},\boldsymbol{h},y_1) + (1-\beta_2)\mathcal{L}({\boldsymbol{\tilde{q}}},\boldsymbol{h},y_2).
\end{align}

\setlength{\tabcolsep}{2mm}
\begin{table}[t]
\footnotesize
  \caption{TDNN encoder architecture used as a front-end model in our experiment. The normal cross-entropy based on speaker labels was used to train this model. }
  \label{table:tdnn}
  \centering
  \vspace{0mm}
  \begin{tabular}{l c c c}
    \toprule
    \multicolumn{1}{l}{\textbf{Layer}} & \multicolumn{1}{c}{\textbf{Layer context}} & \multicolumn{1}{c}{\textbf{Total contexts}} & \multicolumn{1}{c}{\textbf{Input $\times$ output}}\\
    \midrule
    TDNN-ReLU1           & [-2,+2]       & 5    & $C$   $\times$ 512                          \\
    TDNN-ReLU2           & \{-2, 0, +2\} & 9    & 1536  $\times$ 512                          \\
    TDNN-ReLU3           & \{-3, 0, +3\} & 15   & 1536  $\times$ 512                          \\
    TDNN-ReLU4           & \{0\}         & 15   & 512   $\times$ 512                          \\
    TDNN-ReLU5           & \{0\}         & 15   & 512   $\times$ 1500                         \\
    SP              & [0, T)        & $T$  & 1500$T$ $\times$ 3000                         \\
    FC1             & \{0\}         & $T$  & 3000  $\times$ 512                          \\
    FC2             & \{0\}         & $T$  & 512   $\times$ 512                          \\
    Softmax         & \{0\}         & $T$  & 512   $\times$ $M$                            \\
    \bottomrule
  \end{tabular}
  \vspace{-2mm}
\end{table}

\setlength{\tabcolsep}{2.5mm}
\begin{table}[t]
\footnotesize
  \caption{TDNN-ASP encoder architecture used as a front-end model in our experiment. The normal cross-entropy based on speaker labels was used to train this model. }
  \label{table:tdnn-asp}
  \centering
  \vspace{0mm}
  \begin{tabular}{l c c c}
    \toprule
    \multicolumn{1}{l}{\textbf{Layer}} & \multicolumn{1}{c}{\textbf{Layer context}} & \multicolumn{1}{c}{\textbf{Total contexts}} & \multicolumn{1}{c}{\textbf{Input $\times$ output}}\\
    \midrule
    TDNN-ReLU1           & [-2,+2]       & 5    & $C$   $\times$ 512                          \\
    TDNN-ReLU2           & \{-2, 0, +2\} & 9    & 1536  $\times$ 512                          \\
    TDNN-ReLU3           & \{-3, 0, +3\} & 15   & 1536  $\times$ 512                          \\
    TDNN-ReLU4           & \{0\}         & 15   & 512   $\times$ 512                          \\
    TDNN-ReLU5           & \{0\}         & 15   & 512   $\times$ 1500                         \\
    ASP             & [0, T)        & $T$  & 1500$T$ $\times$ 3000                         \\
    FC1             & \{0\}         & $T$  & 3000  $\times$ 512                          \\
    FC2             & \{0\}         & $T$  & 512   $\times$ 512                          \\
    Softmax         & \{0\}         & $T$  & 512   $\times$ $M$                            \\
    \bottomrule
  \end{tabular}
  \vspace{-2mm}
\end{table}

\setlength{\tabcolsep}{2mm}
\begin{table}[t]
\footnotesize
  \caption{ECAPA-TDNN encoder architecture used as a front-end model in our experiment. The normal cross-entropy based on speaker labels was used to train this model. }
  \label{table:ecapa-tdnn}
  \centering
  \vspace{0mm}
  \begin{tabular}{l c c c}
    \toprule
    \multicolumn{1}{l}{\textbf{Layer}} & \multicolumn{1}{c}{\textbf{Layer context}} & \multicolumn{1}{c}{\textbf{Total contexts}} & \multicolumn{1}{c}{\textbf{Input $\times$ output}}\\
    \midrule
    TDNN-ReLU1                  & [-2,+2]       & 5    & $C$   $\times$ 512                     \\
    SE-Res2TDNN1           & \{-2, 0, +2\} & 9    & 1536  $\times$ 512                          \\
    SE-Res2TDNN2           & \{-3, 0, +3\} & 15   & 1536  $\times$ 512                          \\
    SE-Res2TDNN3           & \{-4, 0, +4\} & 24   & 1536  $\times$ 512                          \\
    Concat                 & -             & -    & 512   $\times$ 1536                         \\
    TDNN-ReLU2                  & \{0\}         & 24   & 1536  $\times$ 1536                    \\
    ASP                    & [0, T)        & $T$  & 1536$T$ $\times$ 3072                         \\
    FC1                    & \{0\}         & $T$  & 3072  $\times$ 512                          \\
    FC2                    & \{0\}         & $T$  & 512   $\times$ 512                          \\
    Softmax                & \{0\}         & $T$  & 512   $\times$ $M$                            \\
    \bottomrule
  \end{tabular}
  \vspace{-2mm}
\end{table}

\setlength{\tabcolsep}{3mm}
\begin{table}[t]
\footnotesize
  \caption{SEResNet34 architecture used as a front-end model in our experiment. AM-softmax loss function was used to train this model. }
  \label{table:se-resnet34}
  \centering
  \vspace{0mm}
  \begin{tabular}{l c c c}
    \toprule
    \multicolumn{1}{l}{\textbf{Layer}} & \multicolumn{1}{c}{\textbf{Kernel size}} & \multicolumn{1}{c}{\textbf{Stride}} & \multicolumn{1}{c}{\textbf{Output}}\\
    \midrule
    Conv1           & 3 $\times$ 3 $\times$ 32       & 1 $\times$ 1  & $L$ $\times$ 64 $\times$ 32                          \\
    SE-Res1            & 3 $\times$ 3 $\times$ 32       & 1 $\times$ 1  & $L$ $\times$ 64 $\times$ 32                           \\
    SE-Res2            & 3 $\times$ 3 $\times$ 64       & 2 $\times$ 2  & $L/2$ $\times$ 32 $\times$ 64                           \\
    SE-Res3            & 3 $\times$ 3 $\times$ 128      & 2 $\times$ 2  & $L/4$ $\times$ 64 $\times$ 128                           \\
    SE-Res4            & 3 $\times$ 3 $\times$ 256      & 2 $\times$ 2  & $L/8$ $\times$ 128 $\times$ 256                          \\
    Flatten         & -                              & -             & $L/8$ $\times$ 2048                    \\
    ASP             & -                              & -             & 4096                         \\
    FC              & 512                            & -             & 512                          \\
    AM-softmax      & $N$                            & -             & $N$                            \\
    \bottomrule
  \end{tabular}
  \vspace{-4mm}
\end{table}

\section{Experiments}
\label{section:exp}

\subsection{Datasets}
We conducted all experiments on the CNCeleb \cite{cnceleb1, cnceleb2} dataset, collected from Chinese celebrities in 11 different genres. It contains two subsets called CNCeleb1 \cite{cnceleb1} and CNCeleb2 \cite{cnceleb2}. CNCeleb1 has 800 speakers for training and 200 for evaluation, and CNCeleb2 has 2,000 speakers for training. The dataset includes more than 600,000 utterances in total, and the duration is more than 1,000 hours. In contrast to the single-genre data of the widely used VoxCeleb dataset \cite{voxceleb1,voxceleb2}, CNCeleb contains 11 genres that span interviews, singing, conversation, speech, and so on. The mismatch of genre between enrollment utterance and testing utterance leads to performance degradation of successful speaker verification models such as x-vector and ResNet with margin-based softmax on the VoxCeleb dataset \cite{cnceleb2}. In our experimental configuration, training data from CNCeleb1 and CNCeleb2 were used for training, and the performance evaluation was conducted on the CNCeleb1 test dataset.

In addition to the training and evaluation datasets mentioned above, we use the MUSAN \cite{musan} and RIRs \cite{rirs} datasets for noise augmentation. The former contains three types of noise, and the latter contains reverberation data in several different conditions. 

\subsection{Speaker Encoders}
\label{section:baseline}
In our experiments, we trained four different but representative speaker encoders with 23-dimensional Mel-Frequency Cepstrum Coefficients (MFCCs) as speaker embedding extractors. In the training stage of the speaker encoder, noise augmentation with MUSAN, RIRs, mixup, and SpecAug \cite{specaug} are utilized to augment the training data. For each speaker encoder, we utilized cosine similarity, PLDA, NPLDA, and attention back-end with the AGE2E loss and the focal loss on the corresponding speaker embeddings as baseline systems.

\subsubsection{TDNN}
The first speaker encoder is TDNN, the detailed configuration of which is provided in Table \ref{table:tdnn}. This speaker encoder uses TDNN as a frame-level feature extractor, simple statistics pooling for aggregation, and a plain softmax loss function, without using any attention mechanism. 

\subsubsection{TDNN-ASP}
The TDNN-ASP speaker encoder, whose architecture is shown in Table \ref{table:tdnn-asp}, is similar to the above TDNN model except for the aggregation method. It uses attentive statistics pooling \cite{xvector-asp}, which summarizes frames using different weights generated by the attention mechanism.

\subsubsection{ECAPA-TDNN}
ECAPA-TDNN incorporates SE channel attention into the frame-level feature extractor and concatenates shallow and deep feature maps before the attention-based aggregation layer. Note that \cite{ecapa-tdnn} used only one FC layer after aggregation and utilized additive angular margin softmax (AAM-softmax) as the loss function, but in our baseline system, we follow the architecture of x-vector in \cite{xvector1}, which used two FC layers after aggregation and plain cross-entropy as the loss function. The details are provided in Table \ref{table:ecapa-tdnn}.

\subsubsection{SE-ResNet34}
The final speaker encoder is SE-ResNet34, whose architecture is shown in Table \ref{table:se-resnet34}. It embeds SE channel attention in a ResNet34 model to make use of global information. It also uses AM-softmax as the loss function to maximize the between-class distance and minimize within-class variance.

\setlength{\tabcolsep}{0.4mm}
\begin{table}[t]
\footnotesize
  \caption{Comparison of fully E2E method with separated speaker encoder and back-ends. ``Concat" represents concatenating operation. ``Mean" represents averaging operation.}
  \label{tab:exp1}
  \vspace{0mm}
  \centering
  \begin{tabular}{l c c c r r}
    \toprule
    \multicolumn{1}{c}{\textbf{Speaker encoder}} & \multicolumn{1}{c}{\textbf{Back-end}} & \multicolumn{1}{c}{\textbf{E2E}} & \multicolumn{1}{c}{\textbf{Enroll process.}} & \multicolumn{1}{c}{\textbf{EER(\%)}} & \multicolumn{1}{c}{\textbf{minDCF(0.01)}}\\
    \midrule
    TDNN        & Cosine         & $\times$     & Concat         & 15.25           & 0.6784                \\
    TDNN        & Cosine         & $\times$     & Mean           & 14.03           & 0.6700                \\
    TDNN \cite{cnceleb2}  & PLDA & $\times$     & Concat         & 12.52           & \textendash                     \\
    TDNN        & PLDA           & $\times$     & Concat         & 12.09           & 0.6105                \\
    TDNN        & PLDA           & $\times$     & Mean           & 13.29           & 0.6522                \\
    TDNN        & NPLDA          & $\times$     & Concat         & 11.11           & 0.5274                \\
    TDNN        & NPLDA          & $\times$     & Mean           & 10.63           & 0.5335                \\
    TDNN        & Attention      & $\times$     & Attention      & 9.49            & 0.5405                \\ 
    TDNN        & Attention      & $\checkmark$ & Attention      & \textbf{8.89}   & \textbf{0.5207} \\ \midrule
    TDNN-ASP    & Cosine         & $\times$     & Concat         & 13.56           & 0.6460          \\
    TDNN-ASP    & Cosine         & $\times$     & Mean           & 11.63           & 0.6157          \\
    TDNN-ASP    & PLDA           & $\times$     & Concat         & 11.63           & 0.6157          \\
    TDNN-ASP    & PLDA           & $\times$     & Mean           & 10.67           & 0.5990          \\
    TDNN-ASP    & NPLDA          & $\times$     & Concat         & 10.90           & 0.5322          \\
    TDNN-ASP    & NPLDA          & $\times$     & Mean           & 10.46           & 0.5414          \\
    TDNN-ASP    & Attention      & $\times$     & Attention      & 9.54            & 0.5447          \\
    TDNN-ASP    & Attention      & $\checkmark$ & Attention      & \textbf{9.02}   & \textbf{0.5238} \\ \midrule
    ECAPA-TDNN  & Cosine         & $\times$     & Concat         & 13.19           & 0.6270          \\
    ECAPA-TDNN  & Cosine         & $\times$     & Mean           & 11.79           & 0.6117          \\
    ECAPA-TDNN  & PLDA           & $\times$     & Concat         & 10.54           & 0.5806          \\
    ECAPA-TDNN  & PLDA           & $\times$     & Mean           & 9.51            & 0.5665          \\
    ECAPA-TDNN  & NPLDA          & $\times$     & Concat         & 10.09           & 0.5186          \\
    ECAPA-TDNN  & NPLDA          & $\times$     & Mean           & 9.70            & 0.5186          \\
    ECAPA-TDNN  & Attention      & $\times$     & Attention      & 8.63            & 0.4958          \\ 
    ECAPA-TDNN  & Attention      & $\checkmark$ & Attention      & \textbf{8.04}   & \textbf{0.4879} \\ \midrule
    SE-ResNet34 & Cosine         & $\times$     & Concat         & 11.23           & 0.5903              \\
    SE-ResNet34 & Cosine         & $\times$     & Mean           & 10.26           & 0.5687              \\
    SE-ResNet34 & PLDA           & $\times$     & Concat         & 10.81           & 0.6078              \\
    SE-ResNet34 & PLDA           & $\times$     & Mean           & 10.48           & 0.6006              \\
    SE-ResNet34 & NPLDA          & $\times$     & Concat         & 49.17           & 1.0000              \\
    SE-ResNet34 & NPLDA          & $\times$     & Mean           & 53.91           & 1.0000              \\
    SE-ResNet34 & Attention      & $\times$     & Attention      & 9.14            & 0.5414              \\
    SE-ResNet34 & Attention      & $\checkmark$ & Attention      & \textbf{8.62}   & \textbf{0.5229}     \\
    \bottomrule
  \end{tabular}
  \vspace{-4mm}
\end{table}

\subsection{Training Methodology}
Since we conducted several experiments on the speaker encoder, back-end model, and fully E2E model separately, in this section, we describe the training methodology for each of these models individually.

For the speaker encoder, we used the above models to extract speaker embeddings for all back-end models. TDNN series were trained using the AdamW optimizer \cite{AdamW} for 21 epochs with a learning rate of 0.001, a momentum of 0.99, and a second-order momentum of 0.999. We used cosine annealing with a restart strategy \cite{cosine_with_warm_restarts} to schedule the learning rate with three epochs and two factors. The SE-ResNet34 was trained using a simple SGD optimizer for 12 epochs with a learning rate of 0.01, a momentum of 0.9, and a  learning rate scheduler that multiplies the learning rate by 0.1 if no improvement on the validation loss is seen for 10,000 iterations.\footnote{It is implemented using the \texttt{ReduceOnPlateau} Pytorch API \cite{pytorch}.} All speaker encoders were trained using the ASV-Subtools \cite{asv-subtools} toolkit.

Four back-end models were compared in our experiments. Three of these were trainable ones, namely, PLDA, NPLDA, and attention back-end; cosine similarity was also used. For PLDA, we followed the implementation of KALDI \cite{kaldi} in which two-covariance PLDA training follows mean subtraction for centralization and linear discriminant analysis (LDA) for dimension reduction from 512 to 256. The PLDA model was trained using all speaker embeddings extracted from the training data (including the augmented data) for ten iterations using the EM algorithm. As for the NPLDA model, we used the configuration of \cite{nplda} and all speaker embeddings from the training data to train it. Parameters of the previous PLDA model were used to initialize the NPLDA model. We trained the attention back-end by using the same configuration as \cite{attention-backend} for 40 epochs.

In order to compare other fully E2E methods with our proposed joint approach, we also trained three baseline systems: the tuple E2E (TE2E) \cite{google-e2e}, generalized E2E (GE2E) \cite{google-ge2e} and NPLDA E2E \cite{nplda-e2e} models. Both the TE2E and GE2E models were trained using an SGD optimizer for 50 epochs. Its learning rate was initialized at 0.01 and decayed by 0.1 every 15 epochs. As for the NPLDA E2E model, we followed the training configuration described in \cite{nplda-e2e}. The previously trained speaker encoder and NPLDA back-end were used to initialize the NPLDA E2E model for better optimization. For training our joint model, the parameters of the pretrained speaker encoder were used to initialize the speaker encoder component, and the back-end component was initialized randomly. The model was trained using an SGD optimizer with a learning rate initialized at 0.0001 and decayed by 0.95 every epoch for 50 epochs.

\subsection{Comparison of Fully E2E Method with Separated Speaker Encoders and Back-ends}
In the first experiment, in order to determine the effect of E2E joint optimization, all the mentioned speaker encoders and back-end models discussed above were trained, and then compared with the proposed fully E2E models by jointly optimizing the combined speaker encoders with attention back-end models. The results of EER and minDCF(0.01) are divided into four groups in accordance with the type of speaker encoders, as shown in Table \ref{tab:exp1}. As we can see, for all the speaker encoders we used, jointly optimizing the pretrained speaker encoder with the attention back-end model outperforms the corresponding separated speaker encoders with all back-end models. These results also demonstrate that our previously proposed attention back-end in \cite{attention-backend} surpasses other conventional back-end models and the neural network-based NPLDA in the case of multiple enrollment utterances.

Additionally, no matter which speaker encoder or which back-end model we used, the mean operation of averaging speaker embeddings on multiple enrollments is better than the operation of concatenating waveforms of various enrollments. Therefore, in the following experiments, apart from the attention mechanism for the enrollment process and the proposed fully E2E method, all other reported results are based on the mean operation for multiple enrollments.

\subsection{Comparison with Other E2E methods}
Various previously developed E2E methods \cite{google-e2e, google-ge2e, nplda-e2e} have also reported excellent results for the speaker verification task. In this part, we compared our proposed fully E2E method with these SOTA E2E methods based on the ECAPA-TDNN speaker encoder, as shown as Table \ref{tab:comparison_e2e}. The ECAPA-TDNN speaker encoder was first trained based on a speaker classification task and then fine-tuned with different E2E methods. The GE2E method is an improved version of the TE2E method, and as we can see in the table, GE2E is much better than TE2E in terms of the EER and minDCF(0.01) metrics. Our proposed fully E2E method outperformed both the GE2E and NPLDA-E2E methods in terms of the EER metric, by relatively 21.4\% and 13.0\%, respectively, which indicates that the attention mechanism in our method is much more suitable for handling the case of multiple enrollments. As for the minDCF(0.01) metric, our fully E2E method is slightly better than NPLDA-E2E, which is far better than the GE2E method.

\setlength{\tabcolsep}{4.5mm}
\begin{table}[t]
\footnotesize
  \caption{Comparison with other E2E methods based on ECAPA-TDNN speaker encoder.}
  \label{tab:comparison_e2e}
  \centering
  \vspace{0mm}
  \begin{tabular}{l r c}
    \toprule
    \multicolumn{1}{l}{\textbf{Method}} & \multicolumn{1}{c}{\textbf{EER(\%)}} & \multicolumn{1}{c}{\textbf{minDCF(0.01)}}  \\ 
    \midrule
    TE2E \cite{google-e2e} & 10.64 & 0.5769 \\
    GE2E \cite{google-ge2e} & 10.23 & 0.5535 \\
    NPLDA-E2E-aDCF \cite{nplda-e2e} & 9.24 & 0.4922 \\
    Fully E2E & \textbf{8.04} & \textbf{0.4879} \\
    \bottomrule
  \end{tabular}
   \vspace{-2mm}
\end{table}

\subsection{Effect of Proposed Elements}
There are three novel points in our proposed fully E2E method. To figure out the contribution of each component, we conducted an experiment that integrated each part one by one to make the contribution of each element clear. In addition, the separated ECAPA-TDNN speaker encoder with the conventional PLDA back-end model and the neural network-based attention back-end model are compared, as shown in Table \ref{tab:effect_proposed_points}. From the first and second lines of this table, it is obvious that the attention back-end proposed in \cite{attention-backend} can make better use of multiple enrollments compared with the PLDA back-end no matter which evaluation metrics we used. Comparing the second and third lines, we found that fully E2E fine-tuning can overcome some of the intrinsic biases in the speaker encoder introduced by the classification-based loss function. After applying E2E fine-tuning, the EER metric was improved by 4.8\% relatively, although the minDCF(0.01) metric was slightly worse than the attention back-end. From the fourth line, we found that focal loss is able to dynamically alleviate the problem of unbalanced data, as mentioned in the previous section. Finally, the fifth line proves that embedding-level mixup can improve the performance further by increasing the diversity of training data.

\setlength{\tabcolsep}{2mm}
\begin{table}[t]
\footnotesize
  \caption{Effect of proposed elements based on ECAPA-TDNN speaker encoder.}
  \label{tab:effect_proposed_points}
  \centering
  \vspace{0mm}
  \begin{tabular}{l r c}
    \toprule
    \multicolumn{1}{l}{\textbf{System}} & \multicolumn{1}{c}{\textbf{EER(\%)}} & \multicolumn{1}{c}{\textbf{minDCF(0.01)}}  \\ 
    \midrule
    ECAPA-TDNN/PLDA & 9.51 & 0.5665 \\
    ECAPA-TDNN/Attention backend \cite{attention-backend} & 8.93 & 0.5043 \\
    +E2E fine-tune & 8.50 & 0.5075 \\
    ++Focal loss & 8.26 & 0.4909 \\
    +++Embedding mixup & 8.04 & 0.4879\\
    \bottomrule
  \end{tabular}
   \vspace{-4mm}
\end{table}

\subsection{Effect of Number of Enrollment Utterances}
\begin{figure}[!t]
\centering
\includegraphics[width=3.4in]{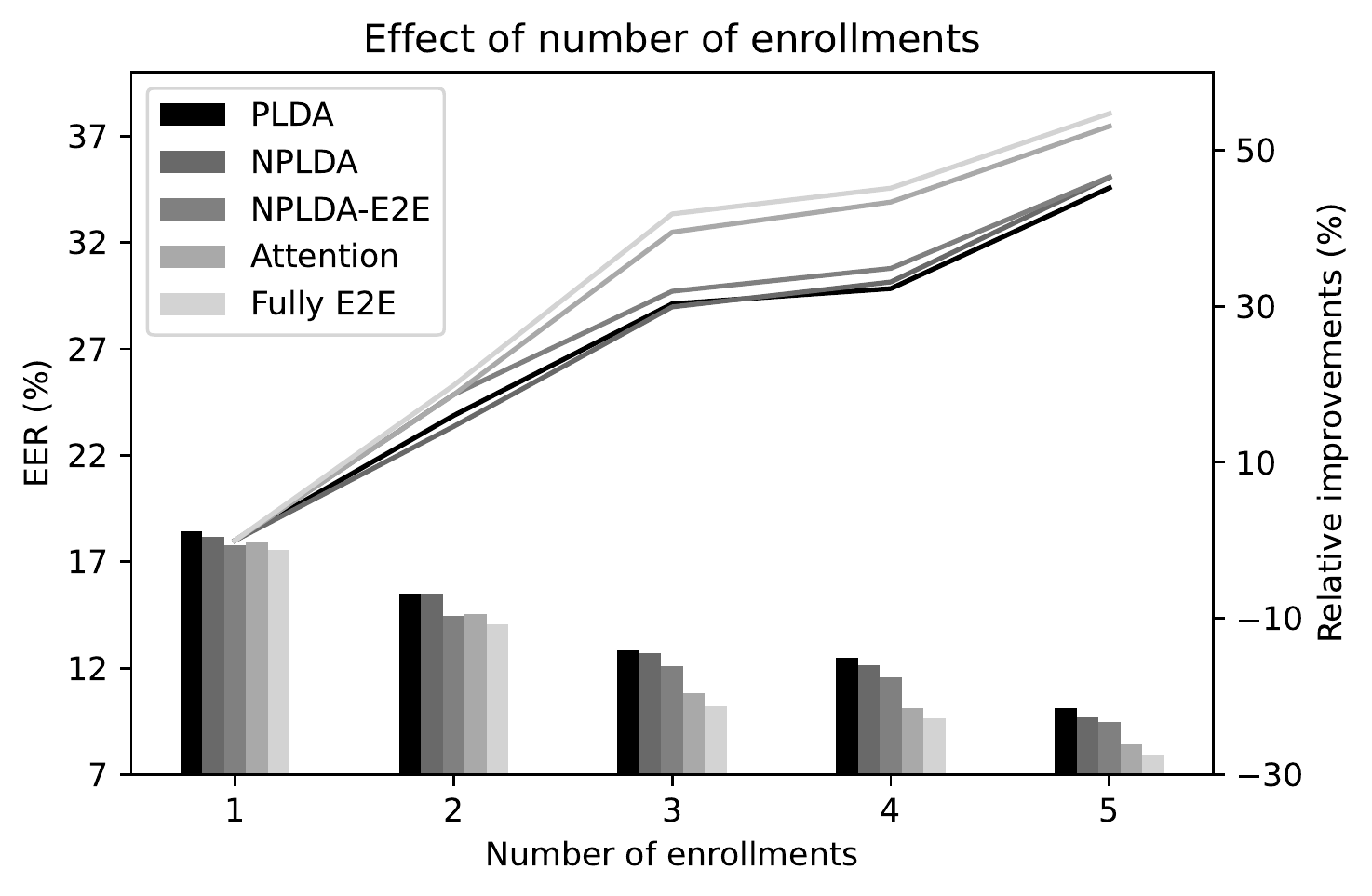}
\caption{Effect of the number of enrollments. The horizontal axis represents the number of enrollments, where 5 means equal to or higher than 5. The left vertical axis shows EER in percentage. The right vertical axis shows the relative improvements in percentage compared with the single enrollment case for each system.}
\label{fig:impact}
\vspace{-4mm}
\end{figure}

To further investigate the performance of our proposed E2E method, we experimented with varying the number of enrollment utterances made by individual speakers. Instead of using the official CNCeleb evaluation protocol, we selected speakers from this protocol whose numbers of enrollments are equal to or more than five as enrollment speakers for a fair comparison. In this experiment, the number of enrollment utterances is gradually increased from $K=1$ to $K \geq 5$  to examine the effect of the number of enrollments. The results shown in Table \ref{tab:result1-breakdown} indicate that enrollment with multiple utterances is much better than using a single one in all situations, as expected. Moreover, as shown in Fig. \ref{fig:impact}, it is obvious that our fully E2E model not only performs better for multiple enrollment utterances but is also competitive in the case of a single enrollment utterance.

\setlength{\tabcolsep}{0.6mm}
\begin{table}[t]
\footnotesize
  \caption{Breakdown EER(\%) results according to number of enrollment utterances of individual speakers included in the eval set. For enrollment processing, concatenation and mean operation were used for ECAPA-TDNN cases.}
  \label{tab:result1-breakdown}
  \centering
  \vspace{0mm}
  \begin{tabular}{l r r r r r}
    \toprule
    \textbf{System} & \textbf{$K=1$} & \textbf{$K=2$} & \textbf{$K=3$} & \textbf{$K=4$} & \textbf{$K \geq 5$}\\
    \midrule
    ECAPA-TDNN/PLDA     & 18.44 & 15.49 & 12.85 & 12.49 & 10.11         \\
    ECAPA-TDNN/NPLDA & 18.15 & 15.50 & 12.72 & 12.14 & 9.71 \\
    ECAPA-TDNN/NPLDA-E2E & 17.76 & 14.44 & 12.09 & 11.57 & 9.49 \\
    ECAPA-TDNN/Attention backend & 17.90 & 14.56 & 10.83 & 10.14 & 8.40 \\
    Fully E2E & \textbf{17.57} & \textbf{14.08} & \textbf{10.22}  & \textbf{9.64} & \textbf{7.96} \\
    \bottomrule
  \end{tabular}
  \vspace{-4mm}
\end{table}

\section{Conclusion}
\label{section:conclusion}
In this paper, we proposed a fully E2E method based on our previously proposed attention back-end model for ASV in order to handle the case of multiple enrollments. To solve the problem of unbalanced training data and to increase the diversity of this data, we incorporate the focal loss as well as a unique embedding-level mixup strategy to optimize the models in a fully E2E style. Experiments on the CNCeleb dataset showed that the proposed approach achieved relative improvements of 21.4\% and 13.0\% over the two compared SOTA E2E methods, respectively. Moreover, compared with separated ECAPA-TDNN speaker encoder and back-end models, our method outperformed the best baseline by an average of 6.8\% relatively, thus demonstrating the advantages of the proposed fully E2E method. We also conducted an experiment to determine the effect of the number of enrollment utterances and found that our fully E2E method not only performed better for multiple enrollment utterances but was also competitive in the case of a single enrollment utterance.

In future work, we will investigate a new direction considering how to simultaneously solve mismatch problems, including general training and testing mismatch, enrollment and evaluation mismatch, and spoofing attacks. We believe it will make the automatic speaker verification system more practical, especially considering the threat of recent deepfake techniques.


\bibliographystyle{IEEEtran}
\bibliography{main}

\hfill

\end{document}